# A METHODOLOGY TO MANAGE VICTIM COMPONENTS USING CBO MEASURE

N Md Jubair Basha[1] and Salman Abdul Moiz[2]

[1]Assistant Professor,Information Technology, Muffakham Jah College of Engineering & Technology, Hyderabad, INDIA

jubairbasha@mjcollege.ac.in

[2]Professor , Information Technology, MVSR Engineering College, Hyderabad, INDIA

Salman.abdul.moiz@ieee.org

## ABSTRACT

*The current practices of software industry demands development of a software within time and budget which is highly productive. The traditional approach of developing a software from scratch requires considerable amount of effort. To overcome the drawback a reuse drive software development approach is adopted. However there is a dire need for realizing effective software reuse. This paper presents several measures of reusability and presents a methodology of reconfiguring the victim components. The CBO measure helps in identifying the component to be reconfigured. The proposed strategy is simulated using HR portal domain specific component system.*

## 1. INTRODUCTION

Effective software reuse helps in development of quality product within time and budget. This also helps in reducing the high effort needed for testing and maintenance of the software products.

Effective software reuse is still a challenging task. Some of the components may not be reused effectively because of their highly cohesive nature. Such components have to be reconfigured. In this paper several quantitative measures are presented for software reusability. Further an approach is presented to identify the component(s) which is less reused. Later CBO measure is used to identify the reconfigurable components and a methodology is presented for reconfiguring the component(s).

The remaining part of this paper is organized as follows: section-2 presents the related work, section-3 presents basic concepts of software reuse & domain engineering, section-4 describes the component measures which are realized on HR portal system, section -5 presents a methodology of identifying and realizing the reconfigurable component(s)  and section -6 concludes the paper.

## 2. RELATED WORK

Reusability metrics defines an approach to measure reusable components. Several reusability metrics have been proposed in literature which has less emphasis on quantitative metrics. In [10] reusability metrics are based upon four characteristics viz. Self descriptiveness, modularity, portability and platform independence. However their weights are assumed based on assumed value which is qualitative in nature. In [11] a subset of reusability metrics are proposed. Though this approach is more efficient than non-automatable techniques, however the goal is to reuse the components interfaces only. This approach lacks the reuse measures at the design level. Zhongjie Wang et.al [12], proposed that the deficiencies of the components which are not suitable for reuse has to be redesigned. However no such approach to identify such components is presented.

In this paper certain software reusability measures are presented. Later a quantitative strategy is presented to identify the components which are to be reconfigured.

## 3. SOFTWARE REUSE

Software Reuse is the use of available software or to build new software from software knowledge. Reusable assets can be either reusable software or software knowledge. Reusability is a property of a software asset that indicates it's probability of reuse [1]. Software Reuse means the process that use "designed software for reuse" again and again [2]. By software reusing, we can manage complexity of software development, increase product quality and makes faster production in the organization.

Recently, design reuse has become popular with (object-oriented) class libraries, application frameworks, design patterns and along with the source code [3]. Jianli et al. proposed two complementary methods for reusing existing components. Among them one allows component evolution itself, which is achieved with binary class level inheritance across component modules. The other is by defined semantic entity so that they can be assembled at compile time or bind at runtime. Although component containment still is the main reuse model that leads to contribute the software product lines development [4]. Regarding the components much information has to be collected, maintained and processed for the retrieval of the components. Maurizio has proposed a methodology to automatically build a software catalogue the tools for archiving and retrieval of information are presented [5]. Software Reuse can be broadly divided into two categories viz. product reuse and process reuse. The product reuse includes the reuse of a software component and by producing a new component as a result of module integration and construction. The process reuse represents the reuse of legacy component from repository. These components may be either directly reused or may need a minor modification. The modified software component can be archived by versioning these components. The components may be classified and selected depending on the required domain. [6].

### 3.1 Domain Engineering

Software Reuse can be improved by identifying objects and operations for a class of similar systems, i.e. for a particular domain. In the context of software engineering domains are application areas [7].

There are various definitions of what a *domain* is. Czarnecki's defines [8]:" an area of knowledge scoped to maximize the satisfaction of the requirements of stakeholders, which includes concepts and terminology understood by practitioners in the area and the knowledge of how to build (part of) systems in the area".

Domain Engineering is a process in which the reusable component is developed and organized and in which the architecture meeting requirements of the domain is designed [9].

Domain Engineering can be defined by identifying the candidate domains and performing domain analysis and domain implementation which includes both application engineering and component engineering. Domain Analysis is a continuing process of creating and maintaining the reuse infrastructure in a certain domain. The main objective of domain analysis is to make the whole information readily available. The relevant components (if available) has to be extracted from the repository rather than building the new components from the scratch for a particular domain.

Domain Analysis mainly focuses on reusability of analysis and design, but not code. This can be achieved by building common architectures, generic models or specialized languages that additionally improve the software development process in the specific problem area of the domain. A vertical domain is a specific class of systems. A horizontal domain contains general software parts being used across multiple vertical domains. Mathematical functions libraries container classes and UNIX tools are the examples of horizontal reuse. The purpose of domain engineering is to identify objects and operations of a class in a particular problem domain [7].
In the process of domain analysis, each component identified can be categorized as follows.

- General-purpose components: These components can be used in various applications of different domains (horizontal reuse).
- Domain-specific components: They are more specific and can be used in various applications of one domain (vertical reuse).
- Product-specific components: They are very specific and custom-built for a certain application, they are not reusable or only useful to a small extent.

## 3.2 HR Portal Application

The system is designed in such a way that the client can interact with the web tier and business tier and can connect to the Data Access Object(DAO) component. The web-tier component consists of the JSP's and Servlets.The Business tier consists of the EJB's.The DAO's consists of the classes with its objects communicating to the database.The web-tier components are HttpServlet, HRProcessServlet, Login Servlet, InterviewResultServlet and RegistrationServlet classes.The Business-tier components are EmployeeBean, InterviewResultsBean, HRProcessBean are the three stateless bean classes.The DAO components are BaseDAO, EmployeeDAO, InterviewDAO, HRDAO, ProcessDAO classes.

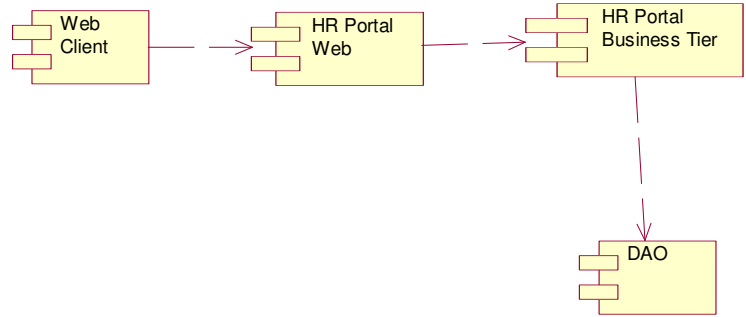

Figure 1. Components of HR Portal Application

## 4. COMPONENT MEASURES

Most of the reuse driven approaches often maintains a repository of reusable components. However an approach is needed to identify those components which are not reused or might have been used very less. Such components are known as Victim components. These components have to modified or reconfigured to achieve high reusability.

The Object Oriented metrics are useful for measuring the effectiveness of component reusability. These metrics are also helpful in identifying the victim components. Some of these metrics are described as follows.

### 4.1 Weighted Component Measure (WCM)

Each component may have several classes and each class may have several methods $m_1, m_2, m_3 .... m_n$. The complexity of all the methods of a particular class is evaluated to find weighted method measure of a class (WMC).

The Weighted Component Measure (WCM) is the sum of Weighted Method Measure of Classes which are composed in the given component.

$$WCM_1 = WMC_1 + WMC_2 + .................... + WMC_n$$

where $WMC_i$ is a Weighted Measure of a classes i=1 to n

The weighted measure of a particular class in a given component is evaluated as

$$WMC_1 = C[M_1] + C[M_2] + C[M_3] + .............. + C[M_K]$$

Where $M_1$ to $M_K$ are the methods of a class.

The complexity of a method is evaluated as follows.

$C[M_j]$= Number of edges – Number of vertices +1   (in a particular flow graph)

or

= Number of Decision Elements of a method + 1

The HR Portal Application consists of three components i.e. Web-Tier, Business-Tier, Data Access Object. These three components consists of 13 classes. Web tier component contains 5 classes, business tier contains 3 classes and DAO component 5 classes. The Weighted Component Measure (WCM) of the Webtier component is as follows

$$WCM_{WBT}= WMC_{HR}+WMC_{INT}+WMC_{RS}+WMC_{LS}$$

The Weighted Measures of a class are as follows:

$WMC_{HR}=C[M_{PR}]+C[M_{R}]$
$WMC_{INT}=C[M_{PR}]+C[M_{AR}]$
$WMC_{RS}=C[M_{PR}]+C[M_{RG}]$
$WMC_{LS}=C[M_{PS}]$

The complexity of the methods is evaluated as follows:

$C[M_{PR}] = 11+1 = 12$
$C[M_{R}] = 10+1 = 11$
$C[M_{PR}] = 11+1 = 12$
$C[M_{AR}] = 6+1 = 7$
$C[M_{PR}] =11+1 =12$
$C[M_{RG}] =8+1= 9$
$C[M_{PS}] = 11+1 =12$

$WMC_{HR}=C[M_{PR}]+C[M_{R}]=12+11=23$
$WMC_{INT}=C[M_{PR}]+C[M_{AR}]=12+7=19$
$WMC_{RS}=C[M_{PR}]+C[M_{RG}]=12+9=21$
$WMC_{LS}=C[M_{PS}]=12$

The Weighted Component Measure(WMC) of the Webtier component is as follows:

$$WCM_{WBT}= WMC_{HR}+WMC_{INT}+WMC_{RS}+WMC_{LS}$$

$WCM_{WBT}=23+19+21+12$
$\quad\quad =75$

The Weighted Component Measure(WCM) of the Businesstier component is as follows:

$$WCM_{BT}= WMC_{EMP}+WMC_{HRPB}+WMC_{IR}$$

The Weighted Measures of a class are as follows:

$WMC_{EMP}=C[M_{GS}]+C[M_{SS}]+ C[M_{A}]+C[M_{ACP}]$
$WMC_{HRPB}=C[M_{RC}]+C[M_{R}]$
$WMC_{IR}=C[M_{VR}]+C[M_{AIR}]$

The cyclomatic complexity of a method is evaluated as follows:

$C[M_{GS}]= 5+1 = 6$
$C[M_{SS}] = 6+1 = 7$
$C[M_{A}] = 7+1 = 8$
$C[M_{ACP}] = 21+1 = 22$
$C[M_{RC}] =12+1 =13$

$C[M_R]$ =10+1= 11
$C[M_{VR}]$ = 9+1 =10
$C[M_{AIR}]$ = 13+1= 14

The Weighted Measures of a class are as follows:

$WMC_{BT=}C[M_{GS}]+C[M_{SS}]+ C[M_A]+C[M_{ACP}]=6+7+8+22=43$
$WMC_{HRPB}=C[M_{RC}]+C[M_R]=13+11=24$
$WMC_{IR}=C[M_{VR}]+C[M_{AIR}]=10+14=24$

The Weighted Component Measure(WCM) of the Businesstier component is as follows:

$WCM_{BT}= WMC_{EMP}+WMC_{HRPB}+WMC_{IR}$

$WCM_{BT}$= 43+24+24

            =91

The Weighted Component Measure(WCM) of the Data Access Object component is as follows:

$WCM_{DAO}= WMC_{BDO}+WMC_{EDO}+WMC_{IDO}+WMC_{HRDO}+WMC_{PDO}$

The Weighted Measures of a class are as follows:

$WMC_{BDO=}C[M_{GC}]+C[M_{CC}]$
$WMC_{HRO}=C[M_{RC}]+C[M_{RE}]$
$WMC_{EDO}=C[M_{GP}]+C[M_{GE}]+ C[M_{AE}]+C[M_{GEE}]+ C[M_{RC}]$
$WMC_{IDO=}C[M_{VIR}]+C[M_{AIR}]$
$WMC_{PDO}=C[M_{RC}]+C[M_{AT}]$

The cyclomatic complexity of a method is evaluated as follows:
$C[M_{GC}]=34+1=35$
$C[M_{CC}]_=4+1=5$
$C[M_{RC}]=12+1=12$
$C[M_{RE}]=13+1=14$
$C[M_{GP}]=9+1=10$
$C[M_{GE}]=13+1=14$
$C[M_{AE}]=21+1=22$
$C[M_{GEE}]=14+1=15$
$C[M_{RC}]=22+1=23$
$C[M_{VIR}]=9+1=10$
$C[M_{AIR}]=13+1=14$
$C[M_{RC}]=29+1=30$
$C[M_{AT}]=6+1=7$

The Weighted Measures of a class are as follows:

$WMC_{BDO=}C[M_{GC}]+C[M_{CC}]=35+5=40$
$WMC_{HRO}=C[M_{RC}]+C[M_{RE}]=13+14=27$
$WMC_{EDO}=C[M_{GP}]+C[M_{GE}]+ C[M_{AE}]+C[M_{GEE}]+ C[M_{RC}]=10+14+22+15+23=84$

$WMC_{IDO} = C[M_{VIR}] + C[M_{AIR}] = 10 + 14 = 24$
$WMC_{PDO} = C[M_{RC}] + C[M_{AT}] = 30 + 7 = 37$

The Weighted Component Measure(WCM) of the Data Access Object component is as follows:

$WMC_{DAO} = WMC_{BDO} + WMC_{EDO} + WMC_{IDO} + WMC_{HRDO} + WMC_{PDO}$

$WCM_{DAO} = 40 + 84 + 24 + 27 + 37 = 212$

## 4.2 Depth of Inheritance Tree Measure

The depth of inheritance is a length from the node where the class is located to the root of the tree. The depth of inheritance tree measures in the maximum length from the node to the root.

The DIT for the Web-tier component is equal to 3.

The DIT for the DAO Component is equal to 2.

The DIT for the business-tier component is equal to 3.

## 4.3 Number of Children Measure

The Number of Children (NOC) measure quantifies the number of immediate subclasses subordinated to a class in the class hierarchy. It measures how many subclasses are inherits the methods of the parent.

The NOC for a class HTTPServlet in component Webtier is 4.

The NOC for a class BaseDAO in DAO component is also 4.

At any particular point of time, if the designer wants to know about which part of the system is not effectively reused then a lookup is to be performed on the component management relation. A Central repository maintains a table for managing a component reuse. This table contains two fields. One specifies the name of the component and the count specifies the number of times the component was reused by several systems.

Table 1. Component Management Relation

| Component | Count of Reuse |
|-----------|----------------|
| Webtier | 12 |
| Businesstier | 5 |
| DAO | 18 |

Table 1. specifies the list of HR portal system components which were reused by several applications. If any component is not used frequently, they are termed as victim components. As Businesstier component was used only 5 times, it can be a candidate of victim component. The victim component has to be reconfigured by dividing it into several parts to increase the reusability count in future. However reconfiguring the non-victim components may also enhance the reusability of the system. The idea is to reconfigure those components which are highly cohesive. Hence, a Coupling Between Object Measure (CBOM) of such components is to be evaluated.

# 5. COMPONENT RECONFIGURATION

Coupling represents the strength of relationship between the components of the system. A well structured system demands loose coupling and a tight cohesion. Those components which are highly cohesive in nature needs to be reconfigured. Coupling between object measure is used to identify the highly cohesive components.

Coupling between object measure (CBOM) for a component is defined as the number of invocations by the specified component. Those components whose CBOM is high or those component(s) of the system whose CBOM is greater than certain scalar value are the components which needs to be reconfigured at the earliest.

Hence a reconfigurable component (Cr) can be identified as

$Cr = Max \{ CBOM(Ci) \}$, where C1, C2,... Cn are components of the system (or)

$Cr_i = \{ CBOM(Ci) > P\}$, where P is a scalar value whose value differs from one domain to the other.

The following figure specifies the number of invocations by different components in HR portal application.

| Hot Spots - Method | Self time [%] ▼ | Self time | Invocations | |
|---|---|---|---|---|
| com.mycompany.hr.dao.BaseDAO.**getConnection** () | ■■■ | 1267 ms (41.4%) | 50 | |
| com.mycompany.hr.dao.EmployeeDAO.**addCandidateProfile** (com.mycompany.hr.vo.CandidateProfile) | ■■ | 946 ms (30.9%) | 20 | |
| com.mycompany.hr.dao.EmployeeDAO.**addEmployeeCredentials** (com.mycompany.hr.vo.EmployeeCredentials) | ■■ | 624 ms (20.4%) | 20 | |
| com.mycompany.hr.dao.EmployeeDAO.**authenticateEmployee** (com.mycompany.hr.vo.EmployeeCredentials) | ▌ | 85.8 ms (2.8%) | 10 | |
| org.apache.jsp.Login_jsp._**jspService** (javax.servlet.http.HttpServletRequest, javax.servlet.http.HttpServlet... | ▌ | 54.6 ms (1.8%) | 25 | |
| com.mycompany.hr.process.__EmployeeBeanRemote_Remote_DynamicStub.**addCandidateProfile** (com.myco... | ▏ | 30.8 ms (1%) | 20 | |
| com.mycompany.hr.process.__EmployeeBeanRemote_Remote_DynamicStub.**authenticate** (com.mycompany.h... | | 15.2 ms (0.5%) | 10 | |
| com.mycompany.hr.servlet.HRProcessServlet.**processRequest** (javax.servlet.http.HttpServletRequest, javax... | | 11.3 ms (0.4%) | 22 | |
| com.mycompany.hr.process.__EmployeeBeanRemote_Remote_DynamicStub.**addCredentials** (com.mycompany... | | 8.7 ms (0.3%) | 20 | |
| org.apache.jsp.AddCandidate_jsp._**jspService** (javax.servlet.http.HttpServletRequest, javax.servlet.http.Htt... | | 5.47 ms (0.2%) | 23 | |
| com.mycompany.hr.servlet.LoginServlet.**processRequest** (javax.servlet.http.HttpServletRequest, javax.servl... | | 3.21 ms (0.1%) | 10 | |
| org.apache.jsp.Welcome_jsp._**jspService** (javax.servlet.http.HttpServletRequest, javax.servlet.http.HttpSer... | | 1.86 ms (0.1%) | 6 | |
| com.mycompany.hr.process.EmployeeBeanBean.**authenticate** (com.mycompany.hr.vo.EmployeeCredentials) | | 1.17 ms (0%) | 10 | |
| org.apache.jsp.ViewProfile_jsp._**jspService** (javax.servlet.http.HttpServletRequest, javax.servlet.http.HttpSe... | | 0.856 ms (0%) | 3 | |
| com.mycompany.hr.servlet.HRProcessServlet.**doGet** (javax.servlet.http.HttpServletRequest, javax.servlet.htt... | | 0.368 ms (0%) | 22 | |
| com.mycompany.hr.dao.BaseDAO.**<init>** () | | 0.235 ms (0%) | 50 | |
| com.mycompany.hr.dao.EmployeeDAO.**<init>** () | | 0.195 ms (0%) | 50 | |
| com.mycompany.hr.process.EmployeeBeanBean.**addCredentials** (com.mycompany.hr.vo.EmployeeCredentials) | | 0.185 ms (0%) | 20 | |
| com.mycompany.hr.servlet.LoginServlet.**doPost** (javax.servlet.http.HttpServletRequest, javax.servlet.http.Htt... | | 0.177 ms (0%) | 10 | |
| com.mycompany.hr.process.EmployeeBeanBean.**addCandidateProfile** (com.mycompany.hr.vo.CandidateProfile) | | 0.175 ms (0%) | 20 | |
| com.mycompany.hr.vo.EmployeeCredentials.**<init>** () | | 0.117 ms (0%) | 60 | |
| com.mycompany.hr.process.__EmployeeBeanRemote_Wrapper.**addCandidateProfile** (com.mycompany.hr.vo... | | 0.114 ms (0%) | 20 | |
| com.mycompany.hr.process.__EmployeeBeanRemote_Wrapper.**addCredentials** (com.mycompany.hr.vo.Employ... | | 0.109 ms (0%) | 20 | |
| com.mycompany.hr.vo.CandidateProfile.**<init>** () | | 0.104 ms (0%) | 40 | |
| com.mycompany.hr.process.__EmployeeBeanRemote_Wrapper.**authenticate** (com.mycompany.hr.vo.Employee... | | 0.068 ms (0%) | 10 | |
| com.mycompany.hr.process.__EmployeeBeanRemote_Wrapper.**<init>** (com.mycompany.hr.process._EmployeeB... | | 0.066 ms (0%) | 2 | |
| com.mycompany.hr.servlet.LoginServlet.**<init>** () | | 0.033 ms (0%) | 1 | |
| com.mycompany.hr.process.__EmployeeBeanRemote_DynamicStub.**<init>** () | | 0.028 ms (0%) | 1 | |
| com.mycompany.hr.servlet.HRProcessServlet.**<init>** () | | 0.026 ms (0%) | 1 | |

Figure 2.Invocations occurred for different components of HR Portal Application

Based on the invocations specified in the above figure, CBOM is evaluated to identify one reconfigurable component.
As there are three components of the HR Portal application viz., Web-tier, Business-tier and DAO, the component with highest CBOM is the candidate for reconfigurable component.

Cr = Max { CBOM(WBR), CBOM(BR), CBOM(DAO) }

Cr = Max { 180, 95,224}= 224

Hence the DAO component has to further reconfigured to reduce the coupling. This is achieved by dividing the DAO component into two sub DAO components viz., $DAO_1$ & $DAO_2$. The division of the DAO component makes the DAO component less cohesive. This is realized later by evaluating the CBO measure of $DAO_1$ & $DAO_2$

## 6. CONCLUSION

Component reusability helps in developing quality product as the component in the repository is successfully tested. Most of the reusability metrics proposed in literature or either qualitative or they realize only interface reusability metrics. In this paper an effort was made to propose reusable quantitative measures. Components which are less used are identified from the repository known as victim components. Further to achieve high reusability, certain components which are highly cohesive in nature have to be reconfigured. An approach using CBOM was proposed to identify such reconfigurable components. In future, strategies to reconfigure the components are to be realized.

## ACKNOWLEDGEMENTS

The work was partly supported by the R & D Cell of Muffakham Jah College of Engineering & Technology, Hyderabad, India. The authors would like to thank to all the people from Industry and Academia for their active support.

## Authors

**N Md Jubair Basha** received his B.Tech. (IT) and M.Tech (IT) from JNTUH,      Hyderabad He is presently working as Assistant Professor in Department of Information Technology, Muffakham Jah College of Engineering and Technology, Hyderabad, India. His research interest includes Software Reusability, Mobile Computing and Cryptography. He is an active member of IEEE and CSI. You can reach him at nawabjubair@gmail.com.

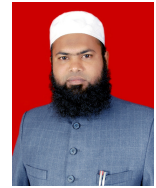

**Dr. Salman Abdul Moiz** is working as a Professor in CSE/IT department at MVSR Engineering College, Hyderabad. He received B.Sc (Electronics) from Osmania University, MCA from Osmania University, M.Tech (CSE) from Osmania University, and M.Phil (CS) from Madurai Kamaraj University and Ph.D (CSE) from Osmania University. He worked as Research Scientist at Centre for Development of Advanced Computing, Bangalore. He has published 31 papers in various National/International Conf erences and Journals. His areas of interests include Mobile databases, Software Process Improvements; Component based software development & Disaster Recovery. He is an active member of IEEE, IETE and CSI.

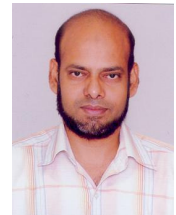